\documentclass[onecolumn, doublespace, 12pt fonts]{aastex6}

\newcommand{\sci}{Sci}

\newcommand{\natco}{NatCo}

\newcommand{\cqgra}{CQGra}

\newcommand{\lrr}{LRR }

\newcommand{\chjaa}{ChJAA}

\begin{document}

\title{New equation of states for postmerger supramassive quark stars}

\author{Ang Li\altaffilmark{1}, Zhen-Yu Zhu\altaffilmark{1}, and Xia Zhou\altaffilmark{2}}

\altaffiltext{1}{Department of Astronomy, Xiamen University, Xiamen, Fujian 361005, China; liang@xmu.edu.cn}
\altaffiltext{2}{Xinjiang Astronomical Observatory, Chinese Academy of Sciences, Urumqi, Xinjiang 830011, China}

\begin{abstract}
Binary neutron star (NS) mergers with their subsequent fast-rotating supramassive magnetars are one attractive interpretation for at least some short gamma-ray bursts (SGRBs), based on the internal plateau commonly observed in the early X-ray afterglow. The rapid decay phase in this scenario signifies the epoch when the star collapses to a black hole after it spins down, and could effectively shed light on the underlying unclear equation of state (EoS) of dense matter. In the present work, we confront the protomagnetar masses of the internal plateau sample from representative EoS models, with the one independently from the observed galactic NS-NS binary, aiming to contribute new compact star EoSs from SGRB observations. For this purpose, we employ various EoSs covering a wide range of maximum mass for both NSs and quark stars (QSs), and in the same time satisfying the recent observational constraints of the two massive pulsars whose masses are precisely measured (around $2 M_{\odot}$). We first illustrate that how well the underlying EoS would reconcile with the current posterior mass distribution, is largely determined by the static maximum mass of that EoS. We then construct 3 new postmerger QS EoSs (PMQS1, PMQS2, PMQS3), respecting fully the observed distribution. We also provide easy-to-use parameterizations for both the EoSs and the corresponding maximum gravitational masses of rotating stars. In addition, we calculate the fractions of postmerger products for each EoS, and discuss potential consequences for the magnetar-powered kilonova model.
\end{abstract}


\section{Introduction}

The mergers of neutron star-neutron star (NS-NS) binaries may result in various remnants, including a prompt black hole (BH), a supramassive or hypermassive millisecond magnetar, and a stable star (SS; being a NS or a quark star, QS). They are in any case highly expected to be associated with multi-band electromagnetic (EM) emission signals coincident with gravitational wave (GW) signals. Particularly, one group of short gamma-ray bursts (SGRBs)~\citep{nsns1,nsns2,nsns3,nsns4} is characterized with extended X-ray internal plateau (IP) emission~\citep{ip00,ip0,ip1,ip2,ip3,ip4,ip5}. Since it is very difficult for a BH engine to power such a plateau, one attractive interpretation is that NS-NS mergers produce a rapidly-spinning, supramassive star (SMS)~\citep{sms1,sms2,sms3,sms4,sms5} rather than a BH, with the rapid decay phase signifying the epoch when the SMS collapses to a BH after the star spins down due to dipole radiation or GW radiation~\citep{collapse4,collapse3,collapse2,collapse1}.

The current modelling of NSs/QSs could qualitatively satisfy the observational constraints of such a SGRB sample~\citep{Lasky2014,ip5}, and requires the postmerger SMS carrying a strong magnetic field $\gtrsim 10^{15}$ G~\citep{collapse1,Li2016b}. The equation of state (EoS) models of QSs have been suggested to be more preferred than those of NSs~\citep{Li2016b}. Consequently, there is an intriguing possibility and new scenario that kilonovae~\citep[e.g.,][]{Li1998} might be associated with postmerger QSs (PMQSs).

Thanks to the development of multi-band observations, including the upcoming new GW detection [Advanced LIGO~\citep{ligo}, Advanced Virgo~\citep{virgo}, and KAGRA~\citep{kagra}], the unknown but important nature of compact stars may be unprecedentedly explored, with the help of the modern EoSs within most-updated nuclear many-body theories~\citep[e.g.,][]{Liapjs}. In the present work, we extend our previous studies of constraining EoSs from SGRB observations, and aim to propose new PMQS EoSs for the use of compact star astronomy, such as SGRB~\citep[e.g.,][]{Lasky2014,ip5,collapse1,Li2016b,Liu2016,Liu2017}, GW~\citep[e.g.,][]{gw}, luminous supernova~\citep[e.g.,][]{sn}, kilonovae~\citep[e.g.,][]{Li1998,kilo}, etc. In addition, we are interested in the relative fractions of postmerger products~\citep{fraction1,fraction2,Sun2017}, and would like study in detail their dependences on a SMS's initial period, as well as the underlying NS/QS EoSs.

The paper is organized as follows. In Section 2, we describe how the observational collapse time of SGRBs are computed in the scenario of magnetar central engine model, and confront the results with various NS/QS EoSs models to those of four observed bursts, respectively, from a IP sample collected by~\citet{Lasky2014}. In Section 3, postmerger product fractions are discussed; New PMQS EoSs are constructed in Section 4. Summary and future perspective are finally presented in Section 5.

\section{Collapse time of the SGRB sample}

Since the rapid decay in X-ray luminosity indicates the spindown-induced collapse of a SMS to a BH, one can combine the standard spin-down formula $P(t)$ with the maximum gravitational mass parameterized as a function of the spin period $M_{\rm max} (P)$, which are written as ($\mu_0 = 1$ in CGS units):
\begin{eqnarray}
\frac{P(t)}{P_0} = \left[ 1 + \frac{4\pi^2B_p^2R^6}{3c^3IP_i^2}t \right]^{1/2}, \\
M_{\rm max} = M_{\rm TOV} (1 + \alpha P^{\beta}).
\end{eqnarray}
$(M_{\rm TOV}, R, I, \alpha, \beta)$ are star parameters calculated from the underlying EoS. $M_{\rm TOV}$, $R$, $I$ are the static gravitational maximum mass, the radius, and the moment of inertia, respectively;  $\alpha, \beta$ are the fitting parameters for $M_{\rm max}$ in Equation~(2). We employ the well-tested $rns$ code (http://www.gravity.phys.uwm.edu/rns) to obtain equilibrium sequences of rapidly rotating, relativistic stars in general relativity. More details about the code can be found in~\citet{rns1,rns2,rns3}. $P_i, B_p$ are, respectively, the initial spin period and the surface dipolar magnetic field.

Setting the protomagnetar mass $M_{\rm p} = M_{\rm max}$ in Equation~(2), we have the collapse time $t_{\rm col}$ defined as a function of $M_{\rm p}$ for each $P = P_i$ in Equation (1):
\begin{eqnarray}
t_{\rm col} = \frac{3c^3I}{4\pi^2B_p^2R^6}\left[ (\frac{M_{\rm p} - M_{\rm TOV}}{\alpha M_{\rm TOV}})^{2/\beta} - P_i^2 \right],
\end{eqnarray}
which is ready to be evaluated with known ($P_i, B_p$). ($P_i, B_p$) can be derived from burst observations assuming electromagnetic dipolar spin-down, with $100\%$ the isotropic efficiency in the conversion between rotational energy and electromagnetic radiation (a reduced efficiency or beam opening angle would lead to a reduction of $P_i$~\citep{Lasky2014}).

We collect those EoS-related parameters $(M_{\rm TOV}, R, P_{\rm K}, M_{\rm max}, I, \alpha, \beta)$ in Table 1, for all nine EoS models used in the present work, including $P_{\rm K}$, the mass-sheding or the Keplerian limit. Nine EoS models include the nonunified models of GM1~\citep{gm1} and APR~\citep{apr}, unified BSk model (BSk20, BSk21)~\citep{bsk}, CDDM model (CIDDM, CDDM1, CDDM2)~\citep{Li2016b}, and MIT$\alpha_s^2$ model (MIT2, MIT3)~\citep{mit}. $\alpha_s^2$ indicates the perturbative QCD corrections to the EoS evaluating up to $O(\alpha_s^2)$ in the strong coupling constant $\alpha_s$~\citep{mit2-3,mit2-2,mit2-1,mit2}. They cover a relatively wide range of maximum mass as well as theoretical models for both NSs and QSs. Among them, a nonunified one (GM1) with the static gravitational maximum mass $M_{\rm TOV} = 2.37 M_{\odot}$ has been widely used and preferred in many previous investigations. We mention here that the nonunified NS models of GM1 and APR are used together with the Negele-Vautherin and the Baym-Pethick-Sutherland crust EoS, as commonly done in the literature.

Following Equation~(3), we then calculate the collapse time of the selected sample of SGRBs (GRB 060801, GRB 080905A, GRB 070724A, GRB 101219A) with IP by Lasky et al. 2014. Figure~1 confronts theoretical results with the corresponding observed values (shown in horizontal lines) for four/five representative NS/QS EoSs. The crosses of the EoSs' results with the corresponding horizontal lines in the figure give the predicted protomagnetar masses for the bursts.

On the other hand, we can necessarily assume the cosmological binary NS mass distribution is the same with the Galactic one, and obtain independently the protomagnetar mass distribution from the observed masses of NSs in binary NS systems: $M = 1.32_{\rm -0.11}^{\rm +0.11} M_{\odot}$, with the errors being the $68\%$ posterior predictive intervals~\citep{2ns}. Based on this and ignoring the ejected material during the merger, one can work out the gravitational mass distribution as: $M = 2.46_{\rm -0.15}^{\rm +0.13} M_{\odot}$~\citep{Lasky2014}, indicated in Figure~1 with shaded regions. An approximate conversion between gravitational mass $M$ and baryonic masse $M_{\rm b}$, $M_{\rm b} = M + 0.075M^2$, has been used to deduce this result~\citep{Lasky2014}.

From Figure~1, it is clear that the $t_{\rm col}$ vs. $M_{\rm p}$ relations are in all cases here nearly vertical curves before crossing, which demonstrates that the required $M_{\rm p}$ related to a burst for each EoS is essentially the EoS's static maximum value: $M_{\rm p} \approx M_{\rm TOV}$. Therefore, we argue that how well the underlying EoS would reconcile with the current posterior mass distribution (with the mean of the distribution $\sim 2.46 M_{\odot}$), is largely determined by the static maximum mass ($M_{\rm TOV}$) of that EoS. Although we only employ the most-preferred ($P_i, B_p$) values for the calculations, the inclusion of observational error intervals in $P_i$ and $B_p$~\citep{Lasky2014}, or using the observed $t_{\rm col}$ only as lower limits~\citep{ip5} would introduce no significant change of this conclusion. The change from NS models (blue curves) to QS models (red curves) also do not affect. One can further notice that in GRB 101219A case, after the crossing, there is a relatively big $M_{\rm p}$ dependence for $t_{\rm col}$, with QS models being more pronounced than NS ones. This can be understood from its small initial period ($P_i \sim 0.95$ ms), which is comparable with the smallest-allowed values $P_{\rm K}$ for these EoSs (see the fifth column of Table 1), and the maximum mass $M_{\rm max}$ in Equation~(2) is a decreasing function with $P$, sharper in the QS case than in the NS case, especially when it is close to $P_{\rm K}$~\citep{Li2016b}.

One related point is that during the deducing of $P_i$ and $B_p$, typical star parameters were usually adopted~\citep[e.g.,][]{ip2,ip4,Lv2017}, such as $M = 1 M_{\odot}$ or $1.4 M_{\odot}, R = 10$ km, $I = 1.5\times10^{45}$ g cm$^2$. This could bring inconsistency to the resulting $P_i$ and $B_p$ when using them together with actually different EoS model (hence different $(M, R, I)$) for further calculations. A more consistent way of deducing could be to employ the corresponding star parameters $(M, R, I)$ from the chosen EoS. One expects it could bring some but nonsignificant changes, since, for example, $P_i \propto R_6^{3/2}$ with $R_6$ the radius in the unit of $10^6$ cm.

\section{Postmerger product fractions}

With the same posterior mass distribution in hand, we can also study the postmerger product fractions, and discuss in detail what a fraction of fast-rotating compact stars would settle to a SMS, which is doomed to collapse to a BH in a range of delay timescales.

For any given initial spin period allowed $P_i~(\leq P_{\rm K}$), the upper bound $M_{\rm sup}$ for the SMS mass is calculated by solving $[(M_{\rm sup} - M_{\rm TOV})/(\alpha M_{\rm TOV})]^{1/\beta} = P_i$ deduced in~\citet{Li2016b}. Setting the lower bound as the nonrotating maximum mass $M_{\rm TOV}$, we can finally evaluate the supramassive NS/QS fraction based on the $M = 2.46_{\rm -0.15}^{\rm +0.13} M_{\odot}$ mass distribution. For the fraction of SS (prompt BH), the two bounds are 0 ($M_{\rm sup}$) and $M_{\rm TOV}$ ($\infty$). Those are done for seven representative EoSs: BSk21, GM1, MIT2, MIT3, CIDDM, CDDM1, CDDM2.

The results are shown in Figure 2. Each EoS line begins with the smallest period allowed (Keplerian limit $P_{\rm K}$) from $rns$ calculations~\citep{rns1,rns2,rns3}. From Panel (a), it is clear that the SS fraction is a constant for each EoS, since it is only determined by the static maximum mass $M_{\rm TOV}$ (as labelled value) together with the observed distribution $M_{\rm p} = 2.46^{+0.13}_{-0.15}$. Consequently, the closer for $M_{\rm TOV}$ to the intermediate value $2.46 M_{\odot}$, the larger the resulting SS fraction. It it the case for both NSs and QSs within different modelling. However, the fractions of SMS (Panel (b)) and BH (Panel (c)) are very sensitive to the initial period, since for them the fast-rotating configurations of the star have to be taken into account, and different NS/QS EoS brings different dependence of the star mass on spin, as have been discussed in detail in our previous work~\citep{Li2016b}. $M_{\rm TOV}$ as well plays a dominating role for the SMS/BH fraction: The closer for $M_{\rm TOV}$ to the intermediate value $2.46 M_{\odot}$, the less stiff for the SMS/BH fraction vs. $P_i$ function. If a typical value $P_i = 1$ ms is used (shown in Figure~(2b) with a vertical dotted line), one can estimate the predicted fraction of SS/SMS/BH based on an EoS model, as done in~\citet{collapse1,Sun2017}.

Recently in the present merger picture, a massive millisecond magnetar model was shown for kilonovae~\citep[namely mergernovae, see e.g.,][]{Yu2013,Yu2015,fraction1}. In observations, some indicative evidences of kilonovae were discovered in the late afterglows of GRBs 130603B \citep{Tanvir2013}, 060614 \citep{Yang2015}, 050704 \citep{Jin2016}. Before \citet{Li1998} has already presented that the radioactive decay in the ejecta originated from a NS-NS merger should power a transient lasting few days. A more detailed description of kilonovae, including the r-process heating due to the disk wind cooling, had been investigated \citep[e.g.,][]{Metzger2010}. Present results in Figure 2 could have important consequences for the ``magnetar-powered mergernova'' scenario. The postmerger product fractions, and more importantly the model of QSs over NSs~\citep{Li2016b}, could be further probed by combined analysis of SGRBs and kilonovae events.

\section{new PMQS EoSs}

This section is devoted to construct new protomagnetar EoSs, based on the conclusion draw above, namely how well the underlying EoS would reconcile with the current posterior mass distribution is largely determined by the static maximum mass ($M_{\rm TOV}$) of that EoS.

We first choose three fixed $M_{\rm TOV}$ values ($2.31 M_{\odot},~2.46 M_{\odot},~2.59 M_{\odot}$) following the distribution with the confidence level of $2\sigma$. Next, considering one of the stiffest unified NS EoS (BSk21~\citep{bsk}) in the market gives $2.28 M_{\odot}$, by reproducing correctly the empirical saturation properties~\citep[e.g.,][]{Li2016c}, we restrict ourselves in the present work to new EoSs for only QSs. NS EoSs could be further softened by one or more types of strangeness phase transitions possible in the NSs' innermost parts, for example, hyperons~\citep[e.g.,][]{Li11y,Li14y}, kaon meson condensation~\citep[e.g.,][]{Li04k,Li06k,Li10k}, $\Delta (1232)$ excitation~\citep[e.g.,][]{Li16d}, quark deconfinement~\citep[e.g.,][]{Li08q,Li15q}. Moreover, to facilitate a EoS model with as little as model parameters, we employ the MIT bag model for constructing the QS EoSs.

The MIT bag model has long been used for the study of QSs following~\citet{mit1-1,mit1-2}, with the usual correction $\sim \alpha_s$ from perturbative QCD. Then the thermodynamic potential of a plasma of massless quarks and gluons was calculated perturbatively to $\sim \alpha_s^2 $ by~\citet{mit2-3,mit2-2} and by~\citet{mit2-1}. The $O(\alpha_s^2)$ pressure for three massless flavors was shown by~\citet{mit2} to be approximated in a similar simple form with the original bag model:
\begin{eqnarray}
P_{\alpha_s^2} = -\Omega_{\alpha_s^2} \approx \frac{3}{4\pi^2}(1 - a_4)(\frac{\mu_{\rm b}}{3})^4 - B_{\rm eff},
\end{eqnarray}
where $\mu_{\rm b}$ is the baryon chemical potential. $a_4 = 1$ corresponds to no QCD corrections. Both $B_{\rm eff}$ and $a_4$ are effective parameters including non-perturbative effects of the strong interactions. They found $a_4$ close to 0.63 with different choice of the renormalization scale, but a relatively large variation for $B_{\rm eff}$. Following~\citet{mit2-qs05,mit2-qs11,mit}, we treat $B_{\rm eff}$ and $a_4$ as free parameters for the study of strongly-coupled dense QS matter, keeping the $a_4$ value at the order of $0.6$.

The QS matter is as usual described as a mixture of quarks ($u,~d,~s$) and electrons, allowing for the transformations mediated by the weak interactions between quarks and leptons:
\begin{eqnarray}
 d \rightarrow u + e + \tilde{\nu}_e,~~u + e \rightarrow d + \nu_e;~~~
 s \rightarrow u + e + \tilde{\nu}_e,~~u + e \rightarrow s + \nu_e;~~~
 s + u \leftrightarrow d + u. \nonumber \\
\end{eqnarray}
The equilibrium composition of QS matter is determined from the $\beta$-stable condition and the electric charge neutrality condition:
\begin{eqnarray}
&& \mu_s = \mu_d = \mu_u + \mu_e,\\
&& \frac{2}{3}n_u - \frac{1}{3}n_d -\frac{1}{3}n_s - n_e =0,
\end{eqnarray}
where the baryon chemical potential is $\mu_{\rm b} = \mu_u + \mu_d + \mu_s$ and the total baryon number density is $n_{\rm b} = (n_u + n_d + n_s)/3$. $n_i$ ($i = u, d, s, e$) is calculated as $n_i = -(\partial \Omega / \partial \mu_i)_{\rm V}$, with the grand canonical potential per unit volume written as:
\begin{eqnarray}
\Omega = \sum_{i} \Omega_i^0 + \frac{3}{4\pi^2}(1 - a_4)(\frac{\mu_{\rm b}}{3})^4 + B_{\rm eff},
\end{eqnarray}
where $\Omega_i^0$ is the grand canonical potential for species $i$ described as ideal relativistic Fermi gases. We take $m_u = m_d = 0$, $m_s = 100$ MeV, and $m_e = 0$. The energy density and pressure can be obtained using standard thermodynamical relations:
\begin{eqnarray}
&& \varepsilon = \sum_{i} \Omega_i^0 + \frac{3}{4\pi^2}(1 - a_4)(\frac{\mu_{\rm b}}{3})^4 +  \sum_{i} \mu_i n_i+ B_{\rm eff},\\
&& P = n_{\rm b}^2\frac{\partial}{\partial n_{\rm b}}(\frac{\varepsilon}{n_{\rm b}}).
\end{eqnarray}

For a given EoS $P(\varepsilon)$, the $rns$ code~~\citep{rns1,rns2,rns3} presents uniformly rotating, axisymmetric, equatorially symmetric configurations of a NS/QS. We first perform such calculations for static stars and find the corresponding maximum mass $M_{\rm TOV}$, which exactly match our three chosen $M_{\rm TOV}$ values. The resulting model parameter sets ($B_{\rm eff}, a_4$) are listed in Table 2, together with their zero-pressure density ($n_{\rm s}$) and energy density ($\varepsilon_{\rm s}$). For completeness, one representative case of soft EoS (MIT2) with $M_{\rm TOV} = 2.08 M_{\odot}$~\citep{mit} is also shown. In each case, the $a_4$ value is around $0.6$ as discussed before, indicating that QCD corrections are inevitable for the study of dense QS matter. $n_{\rm s}$ (or $\varepsilon_{\rm s}$) as the QS surface density decreases with the increase of the stellar mass, and can be regarded as characteristic of the EoS's stiffness.

Detailed formalism in the case of massless quarks are simple and instructive. In this case, $\Omega_q^0 = -\mu_q^4/(4\pi^2)$ ($q = u,d,s$), then from Equation~(8),
\begin{eqnarray}
n_q = -(\partial \Omega/\partial \mu_q)_{\rm V}= \frac{\mu_q^3}{\pi^2} - \frac{1}{\pi^2}(1 - a_4)(\frac{\mu_{\rm b}}{3})^3.
\end{eqnarray}
We have
\begin{eqnarray}
n_{\rm b} = \frac{1}{3}(n_u + n_d +n_s) = \frac{1}{3}\left[\sum_q\frac{\mu_q^3}{\pi^2} - \frac{3}{\pi^2}(1- a_4)(\frac{\mu_{\rm b}}{3})^3 \right]
\simeq \frac{a_4}{\pi^2}(\frac{\mu_{\rm b}}{3})^3,
\end{eqnarray}
where the last inequality in the right hand side is because of the approximation $\mu_u \simeq \mu_d \simeq \mu_s \simeq \mu_{\rm b}/3$ employed.
Finally the EoS of Equations (9)and (10) can be deduced as:
\begin{eqnarray}
&& \varepsilon = \biggl(\frac{9\pi^{2/3}}{4}\biggr)\frac{n_{\rm b}^{4/3}}{a_4^{1/3}} + B_ {\rm eff}~,\\
&& P = \frac{1}{3} \biggl(\frac{9\pi^{2/3}}{4}\biggr)\frac{n_{\rm b}^{4/3}}{a_4^{1/3}} - B_{\rm eff} = \frac{1}{3} \biggl( \varepsilon - 4 B_{\rm eff} \biggr)~.
\end{eqnarray}

As has been pointed out by \citet{mit}, the above $P(\varepsilon)$ relation (hence the mass-radius relation) does not depend on $a_4$, the perturbative QCD corrections term. As can be immediately seen, the mass-radius relation in the present finite-mass case ($m_s = 100$ MeV) has negligible dependence on $a_4$, either. To illustrate this, we extend the $1/3$ index of $a_4$ in Equation~(13) to a free parameter $i$, and do parameter fitting for all four EoS models in Table 2. The pressure in each case is calculated from Equation~(10) to fulfill the thermodynamical equilibrium. The resulting $i$ parameters are listed in Table 2, with the corresponding coefficient of determinations $R^2$ signifying the accuracy of the fitting. The better the fitting, the closer the value of $R^2$ to 1. The comparisons of the density/radius vs. mass relations are shown in Figure~3 in the left/right panel, between the tabulated EoSs and the fitted ones. They in all four cases agree considerably well. Any change of $a_4$ at fixed $B_{\rm eff}$ will modify the curves in the left panel, but not those in the right panel. That is, $a_4$ will affect the star's mass/radius as functions of central density, but hardly the mass-radius relation.

Next, employing the three new EoSs, the rotating configurations of QSs are computed up to its Keplerian limit with $rns$. We further parameterize $M_{\rm max}$ using the formula of Equation~(2). The results are shown in Table 3. Then the collapse time study can be done for the SGRB sample with these new QS EoSs. It is shown in Figure~4. It is clear that these new EoSs' results follow the same conclusion of Figure~1. Namely, $M_{\rm TOV}$ value of a EoS actually determines its goodness when reconciling with the observed posterior mass distribution. This justifies again the strategy of constructing new EoSs fully respecting the distribution.

To further strengthen the usefulness of the new EoSs, we plot in Figure~5 the measured gravitational mass for NSs in binary NSs~\citep{2ns,2ns2}, along with the lines indicating $M_{\rm TOV}$ for three new PMQS EoSs and two representative NS EoSs (GM1, BSk21). These lines assume no mass rejection during merger and the total baryonic mass is conserved. We can see that three new stiff QS EoSs comfortably cover all observed data points, and could be used extensively for theoretical and observational studies related to compact star physics. On the other hand, the ejected mass could be enhanced by the large asymmetry of the merger system. For a mass ratio $\sim 0.75 - 0.85$, reached in the case of PSR J1518+4909, PSR J1756-2251, PSR J1811-1736, PSR J1829+2456 (four green dots), the ejected mass could be as high as $\sim 0.35 M_{\odot}$, according to the simulation of the coalescence of unmagnetized binary NSs~\citep{ejected}. Then the EoS lines should be shifted upwards accordingly~\citep{gw}.

\section{Summary and future perspective}

In this paper, we have extended our previous work on probing compact star EoS from observations of SGRB remnants, to a selected IP sample with known accurate redshift measurements (hence $P_i$ and $B_p$). In particular, we make use of various contemporary EoSs for both for NSs and QSs, and in the same time covering a relatively wide range of maximum mass as well as theoretical frameworks. Using tabulated EoSs, we compute stationary and equilibrium sequences of rapidly rotating, relativistic stars in general relativity from the well-tested $rns$ code, assuming the matter comprising the star to be a perfect fluid. The collapse time of the sample is then theoretically evaluated in the millisecond-magnetar-central-engine model, and is confronted with the observed one, respectively.

We first demonstrated that how well the underlying EoS would reconcile with the current posterior mass distribution $M = 2.46_{\rm -0.15}^{\rm +0.13} M_{\odot}$, is largely determined by the static maximum mass ($M_{\rm TOV}$) of that EoS, and not influenced by the modification of the star parameters ($M,~R, ~I$) induced by rotation. So it is reasonable to construct new compact star EoSs following the observed posterior mass distribution. Within the $95\%$ mass intervals, we then choose three typical values for $M_{\rm TOV}$ of the new EoSs, $2.31 M_{\odot}$, $2.46 M_{\odot}$, $2.59 M_{\odot}$. Eventually three new PMQS EoSs (PMQS1, PMQS2, PMQS3) are provided in the widely-used MIT model, with their easy-to-use parameterizations. We also do the $rns$ calculates for the new EoSs and fit their fast-rotating configurations in analytic forms.

The other meaningful outcome is concerning the merger outcomes (SS, SMS, BH) and their possible fractions, because those are crucial for the EM transient as well as associated GW signature. Our main finding is that those fractions are subject to both the SMS's initial period and the underlying EoSs. The former one's effect is pronounced, especially around the typical value of 1 ms when the postmerger compact star spins super rapidly. The later one's effect essentially manifest itself by the corresponding static maximum mass ($M_{\rm TOV}$) allowed, being compared with the above-mentioned posterior mass distribution. Both two effects are importantly sensitive, allowing us linking the merger observations closely with the underlying dense matter properties and unraveling the later in the near future. The new QS models for kilonovae mentioned in the present work can be then investigated as well.

For future perspective, useful constraints could be possible when combining GW and EM observations from both coalescing NS-NS binaries and isolated NSs. It would be interesting and meaningful to study whether NS EoSs (including hybrid stars) and QS EoSs could be distinguished from GW emission from elliptically deformed pulsars~\citep{Ligw} and NS-NS binaries~\citep{Limhd}.

\begin{acknowledgements}
We would like to thank Tong Liu, En-ping Zhou, Bing Zhang, and He Gao for valuable discussions; The work was supported by the National Natural Science Foundation of China (Nos. U1431107 and 11373006), and the West Light Foundation of Chinese Academy of Sciences (No. ZD201302).
\end{acknowledgements}

\begin{figure*}
\epsscale{1.0}
\plotone{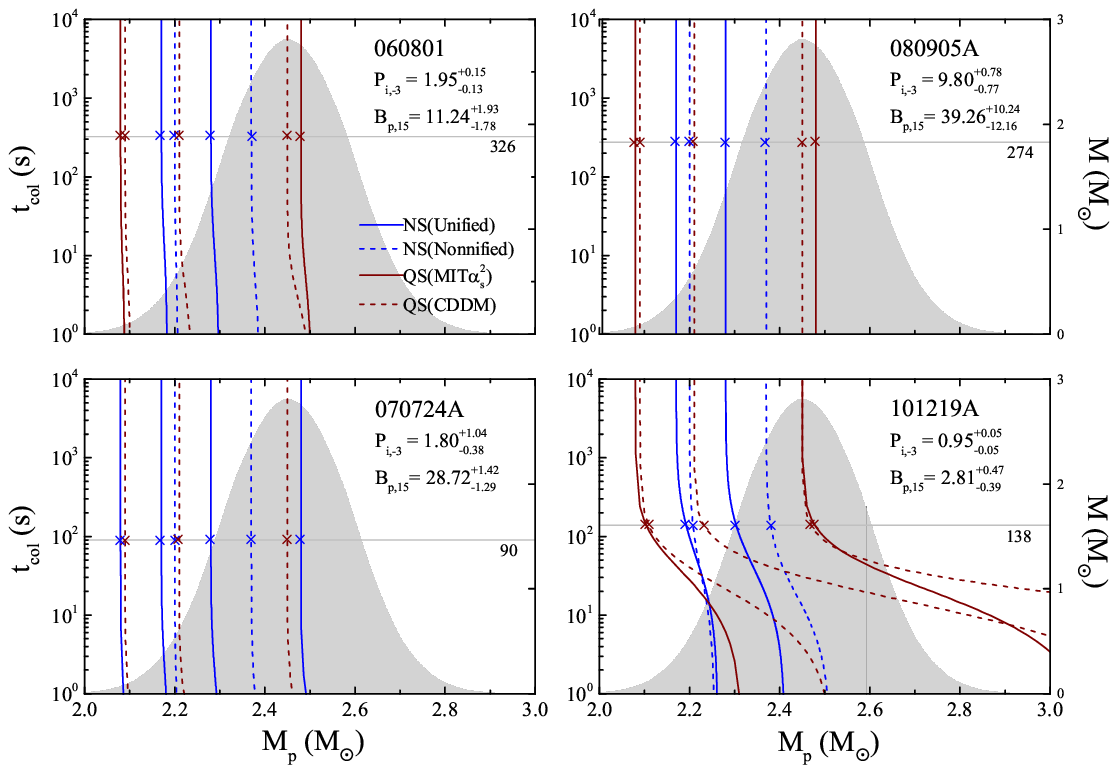}
\caption{Collapse time as a function of the protomagnetar mass for IP sample of~\citet{Lasky2014} (GRB 060801, GRB 080905A, GRB 070724A and GRB 101219A), to be compared with observed value shown in horizontal lines, respectively. The input burst data ($P_i, B_p$) with $68\%$ confidence level are listed as well from~\citet{ip2} and we employ the most-preferred values for calculations. The shaded region is the $M = 2.46_{\rm -0.15}^{\rm +0.13} M_{\odot}$ mass distribution independently derived from the binary NS mass distribution $M = 1.32_{\rm -0.11}^{\rm +0.11} M_{\odot}$~\citep{2ns}. NS (unified) EoSs are BSk20 and BSk21~\citep{bsk}. NS (nonunified) EoSs are GM1~\citep{gm1} and APR~\citep{apr}. QS (MIT) EoSs are MIT2 and MIT3~\citep{mit}. QS (CDDM) are CIDDM, CDDM1, and CDDM2~\citep{Li2016b}. EoSs' details are collected in Table 1.}\label{fig1}
\end{figure*}

\begin{figure*}
\epsscale{1.0}
\plotone{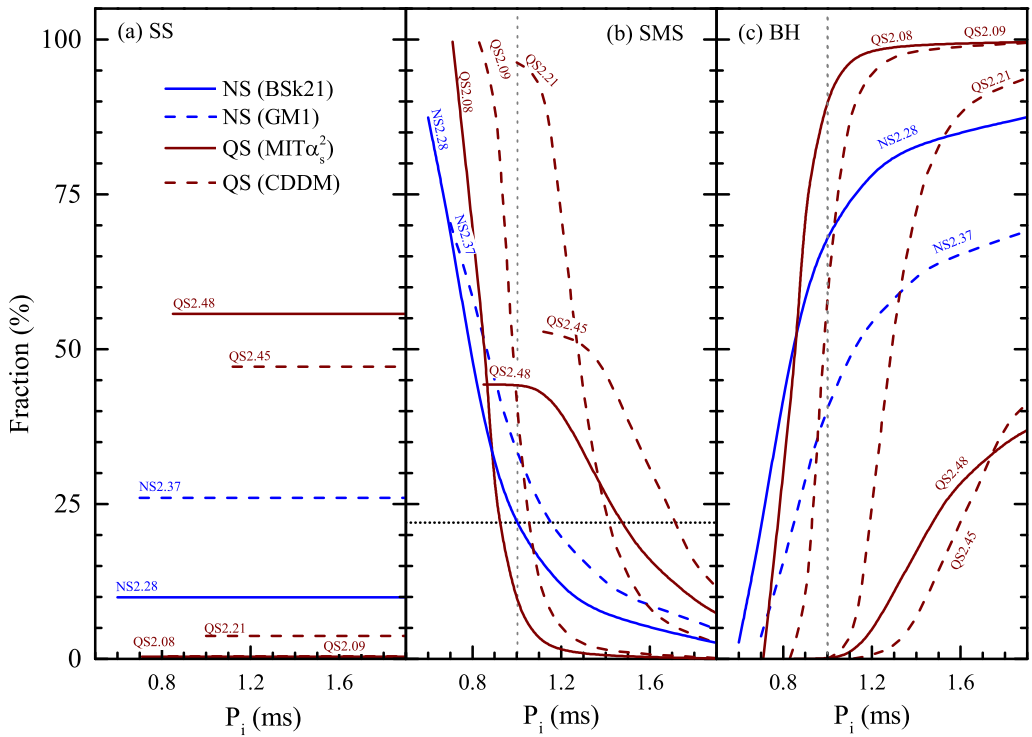}
\caption{Postmerger product fractions for (a) SS, (b) SMS and (c) BH for seven various NS (blue) and QS (red) EoS models, labelled with star type plus the corresponding $M_{\rm TOV}$: Unified BSk21 (blue solid line labelled with NS2.28), nonunified GM1 (blue dashed line labelled with NS2.37), MIT model (red solid lines labelled with QS2.08 (MIT2) and QS2.48 (MIT3)), and CDDM model (red dashed line labelled with QS2.09 (CIDDM), QS2.21(CDDM1) and QS2.45 (CDDM2)). In panel (b) the observed 22\% constraint for SMS from Gao et al. 2016 is shown in horizontal line for comparison. The vertical dotted line in the same panel is for typical initial period of 1 ms. }\label{fig2}
\end{figure*}

\begin{figure*}
\epsscale{1.0}
\plotone{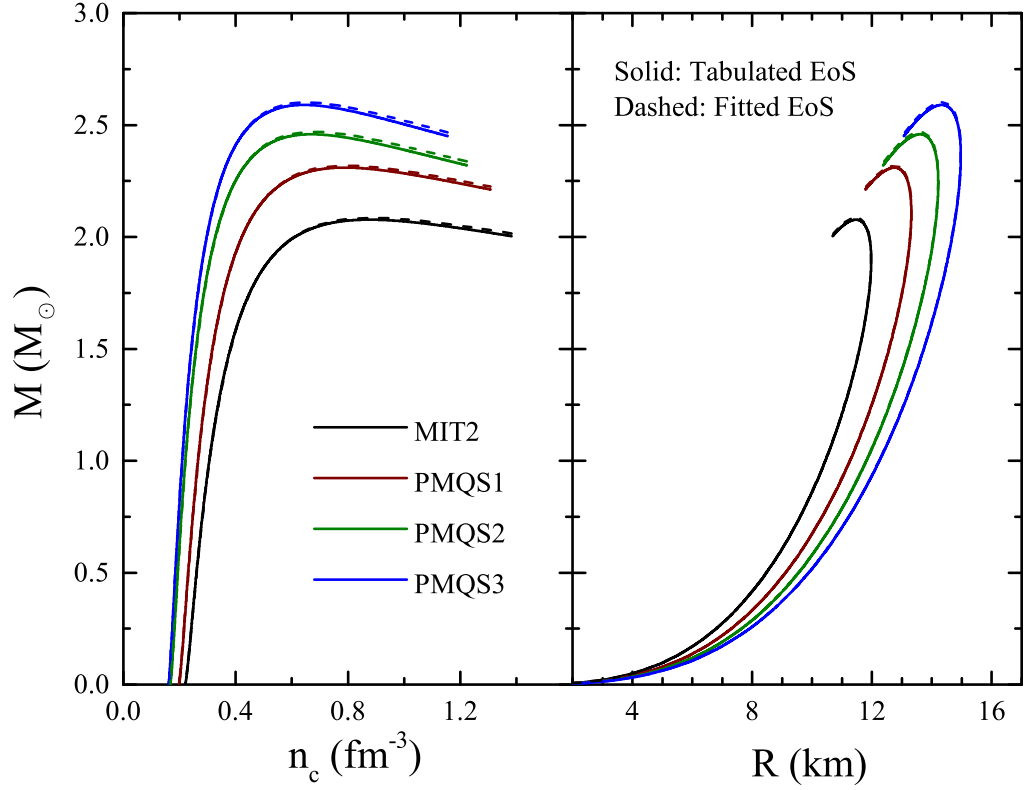}
\caption{(Color online) Comparisons of the density-mass relations (left panel) and mass-radius relations (right panel) between the tabulated EoSs and the fitted ones, for new PMQS EoSs as well as the MIT2 model.}\label{fig3}
\end{figure*}

\begin{figure*}
\epsscale{1.0}
\plotone{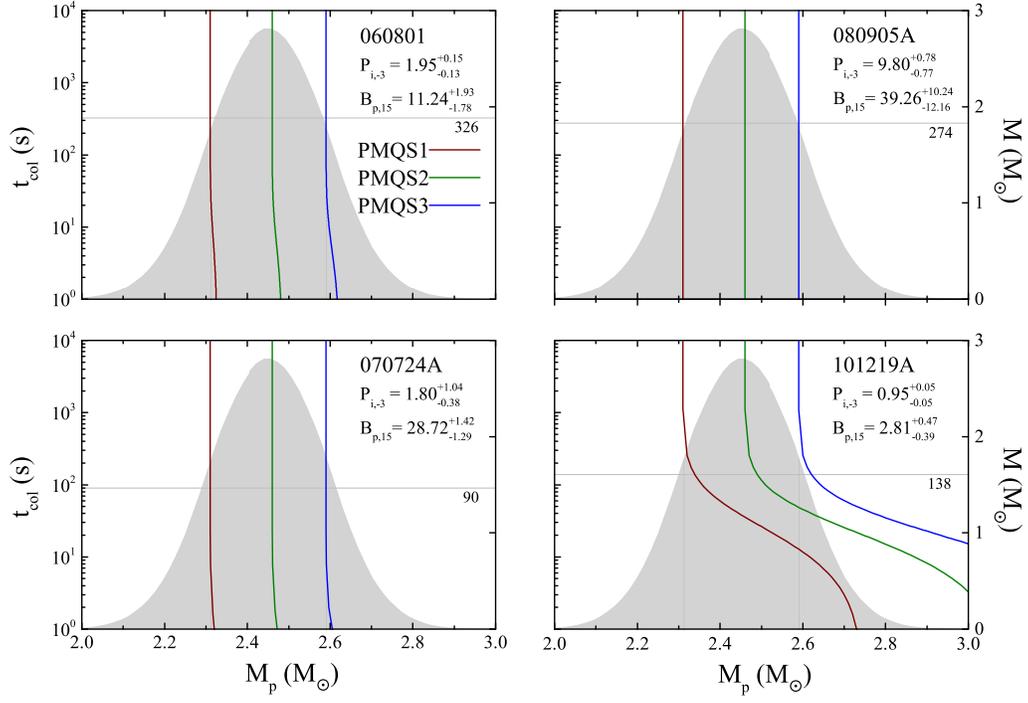}
\caption{(Color online) Same with Figure~1, but with three PMQS EoSs newly-constructed in the present work.}\label{fig4}
\end{figure*}

\begin{figure*}
\epsscale{1.0}
\plotone{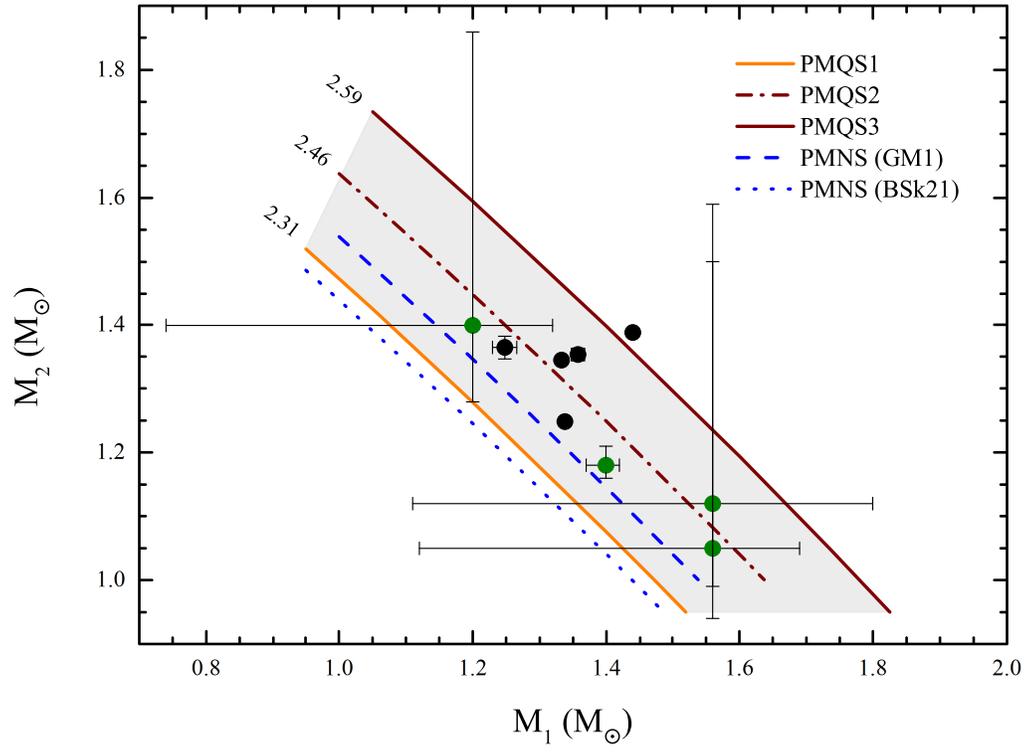}
\caption{(Color online) Measured gravitational mass for NSs in binary NSs~\citep{2ns}, along with lines indicating $M_{\rm TOV}$ for three new PMQS EoSs and two representative NS EoSs (GM1, BSk21).}\label{fig5}
\end{figure*}

\begin{table*}
\tabcolsep 1pt
\caption{NS/QS EoSs investigated in this study: Nonunified GM1, APR, unified BSk model (BSk20, BSk21), CDDM model (CIDDM, CDDM1, CDDM2), and MIT model (MIT2, MIT3). $M_{\rm TOV}$, $R$ are the static gravitational maximum mass and the corresponding radius, respectively; $P_{\rm K}$, $M_{\rm max}$, $I$ are the Keplerian spin limit, the corresponding maximum  mass and maximum moment of inertia, respectively;  $\alpha, \beta$ are the fitting parameters for $M_{\rm max}$ in Equation~(2).}
\vspace*{-12pt}
\begin{center}
\def\temptablewidth{0.96\textwidth}
{\rule{\temptablewidth}{0.5pt}}
\begin{tabular*}{\temptablewidth}{@{\extracolsep{\fill}}ccc|ccc|cccc|ccc}
   \hline
   & && $M_{\rm TOV}$ & $~R~$ && $P_{\rm K}$ &$M_{\rm max}$& $I$ && $\alpha$ & $\beta$ &Ref.  \\
   &EoS && $(M_{\odot})$ & (km)&& ($10^{-3}$s) & $(M_{\odot})$ & $(10^{45} {\rm g~cm}^2)$ && $(10^{-10}P^{-\beta})$  &&\\
   \hline
   &  GM1   && 2.37 & 12.05 && 0.72 & 2.72& 3.33 && 1.58 & -2.84  & ~\citet{Lasky2014}\\
NS &  APR   && 2.20 & 10.0  && 0.52 & 2.61& 2.13 && 0.303 & -2.95 & ~\citet{Lasky2014}\\
   & BSk20  && 2.17 & 10.17 && 0.54 & 2.58& 3.50 && 3.39 & -2.68  & ~\citet{Li2016b} \\
   & BSk21  && 2.28 & 11.08 && 0.60 & 2.73& 4.37 && 2.81 & -2.75  & ~\citet{Li2016b} \\
   \hline
   & CIDDM && 2.09 & 12.43 && 0.83 & 2.93& 8.645 && 2.58/10$^6$ & -4.93 & ~\citet{Li2016b}\\
   & CDDM1 && 2.21 & 13.99 && 1.00 & 3.10& 11.67 && 3.93/10$^6$ & -5.00 & ~\citet{Li2016b}\\
QS & CDDM2 && 2.45 & 15.76 && 1.12 & 3.45& 16.34 && 2.22/10$^6$ & -5.18 & ~\citet{Li2016b}\\
   & MIT2  && 2.08 & 11.48 && 0.71 & 3.00& 7.881 && 1.67/10$^5$ & -4.58 & This work \\
   & MIT3  && 2.48 & 13.71 && 0.85 & 3.58& 13.43 && 3.35/10$^5$ & -4.60 & This work \\
      \hline
\end{tabular*}
      {\rule{\temptablewidth}{0.5pt}}
\end{center}
\end{table*}

\begin{table*}
\tabcolsep 1pt
\caption{Three new stiff QS EoS models ($B_{\rm eff}, a_4$), together with their static maximum mass ($M_{\rm TOV}$), the zero-pressure density ($n_{\rm s}$) and energy density ($\varepsilon_{\rm s}$). Easy-to-use fitting is also done employing the formula extended from the extension of massless-quark case (Equation~(11)). There is one fitting parameter $i$. The coefficient of determinations ($R^2$) are presented as well. One representative case of soft EoS (MIT2) with $M_{\rm TOV} = 2.08 M_{\odot}$ is also shown.}
\vspace*{-12pt}
\begin{center}
\def\temptablewidth{0.96\textwidth}
{\rule{\temptablewidth}{0.5pt}}
\begin{tabular*}{\temptablewidth}{@{\extracolsep{\fill}}cc|cccc|cccc}
   \hline
   && $M_{\rm TOV}$  & $n_{\rm s}$ & $\varepsilon_{\rm s}$ && $B_{\rm eff}^{1/4}$ & $a_4$& $i$ & $R^2$ \\
   && $[M_{\odot}]$  & $[\rm fm^{-3}]$ & $[\rm MeV/fm^{3}]$ && [MeV]           &      &      &       \\
   \hline
    MIT2 && 2.08 & 0.208 & 203.51 && 138 & 0.610 & 0.370670 & 0.999978  \\
    PMQS1&& 2.31 & 0.188 & 164.61 && 131 & 0.754 & 0.393894 & 0.999980  \\
    PMQS2&& 2.46 & 0.157 & 143.49 && 126 & 0.587 & 0.374706 & 0.999973  \\
    PMQS3&& 2.59 & 0.152 & 129.62 && 123 & 0.680 & 0.382617 & 0.999978  \\
      \hline
\end{tabular*}
      {\rule{\temptablewidth}{0.5pt}}
\end{center}
\end{table*}

\begin{table*}
\tabcolsep 1pt
\caption{Same with Table 1, but with three new PMQS EoSs constructed in the present work, and the $R^2$ values for $(\alpha, \beta)$ fitting are presented instead the references.}
\vspace*{-12pt}
\begin{center}
\def\temptablewidth{0.96\textwidth}
{\rule{\temptablewidth}{0.5pt}}
\begin{tabular*}{\temptablewidth}{@{\extracolsep{\fill}}cc|ccc|ccc|ccc}
   \hline
  && $M_{\rm TOV}$ & $~R~$  && $P_{\rm K}$ & $M_{\rm max}$ &    $I$   & $\alpha$  & $\beta$ & $R^2$ \\
  && $(M_{\odot})$ &     (km)        && (ms)        & $(M_{\odot})$ & $(10^{45} {\rm g~cm}^2)$  & $(10^{-15}P^{-\beta})$  & &\\
   \hline
    PMQS1 && 2.31 & 12.75 && 0.79 & 3.34 & 10.83 & 4.39 & -4.51 & 0.98408\\
    PMQS2 && 2.46 & 13.61 && 0.84 & 3.58 & 13.11 & 5.90 & -4.51 & 0.98434\\
    PMQS3 && 2.59 & 14.33 && 0.88 & 3.75 & 15.28 & 9.00 & -4.48 & 0.98412\\
      \hline
\end{tabular*}
      {\rule{\temptablewidth}{0.5pt}}
\end{center}
\end{table*}

\end{document}